Linking Common Multispectral Vegetation Indices to Hyperspectral Mixture Models:
Results from 5 nm, 3 m Airborne Imaging Spectroscopy in a Diverse Agricultural Landscape

Daniel Sousa[1] and Christopher Small[2]

[1]Department of Geography
San Diego State University
San Diego, CA 92181
dan.sousa@sdsu.edu

[2]Lamont-Doherty Earth Observatory
Columbia University
Palisades, NY 10964
csmall@columbia.edu

# Abstract

For decades, agronomists have used satellite and airborne remote sensing to monitor key crop parameters like biomass, fractional cover, and plant health. Vegetation indices (VIs) are popular for this purpose, primarily leveraging the spectral red edge in multispectral imagery. In contrast, spectral mixture models use the full reflectance spectrum to simultaneously estimate area fractions of multiple endmember materials present within a mixed pixel. Mixture modeling is especially useful for hyperspectral imagery because of its ability to discriminate absorption features not resolved by multispectral imagery. Here, we characterize the relationships between hyperspectral endmember fractions and six common multispectral VIs in diverse, globally significant crops and soils of California agriculture. Fractional area of photosynthetic vegetation ($F_v$) was estimated directly from 64,000,000 5 nm, 3-5 m resolution reflectance spectra compiled from a mosaic of 15 AVIRIS-ng flightlines. Simulated Planet SuperDove reflectance spectra were then derived from the AVIRIS-ng spectra, and used to compute six popular VIs (NDVI, NIRv, EVI, EVI2, SR, DVI). Multispectral VIs were compared to hyperspectral $F_v$ using both parametric (Pearson correlation coefficient, ρ) and nonparametric (Mutual Information, MI) similarity metrics. Four VIs (NIRv, DVI, EVI, EVI2) showed strong linear relationships to $F_v$ (ρ > 0.94; MI > 1.2). NIRv and DVI showed strong interrelation (ρ > 0.99, MI > 2.4), but deviated significantly from a 1:1 correspondence to $F_v$. EVI and EVI2 were also strongly interrelated (ρ > 0.99, MI > 2.3) and more closely approximated a 1:1 relationship with $F_v$. In contrast, NDVI and SR showed weaker, nonlinear, heteroskedastic relations to $F_v$ (ρ < 0.84, MI = 0.69). NDVI exhibited especially severe sensitivity to substrate background reflectance (–0.05 < NDVI < +0.6 for unvegetated spectra) and saturation (0.2 < $F_v$ < 0.8 for NDVI = 0.7). These direct observational constraints on multispectral VI and hyperspectral mixture model comparability can serve as a quantitative benchmark for agronomic applications in the coming era of increasing spatial and spectral resolution Earth observation.

# Introduction

Agricultural remote sensing is undergoing a fundamental transformation. On the one hand, imaging spectroscopy (hyperspectral) missions like SBG, EMIT, EnMAP, DESIS, PRISMA, and HISUI promise to improve spectral fidelity *(Thompson et al. 2021; Candela et al. 2016; Iwasaki et al. 2011; Guanter et al. 2015; Green et al. 2020)*, with important agronomic implications *(Liu et al., 2020)*. Simultaneously, CubeSat constellations like Planet enable order-of-magnitude improvements in spatial and temporal resolution of multispectral imagery *(Safyan, 2020)*, allowing for improved precision in crop monitoring – if radiometric inconsistencies can be overcome (*Leach et al., 2019*). The future of spaceborne agronomic applications will require a quantitative understanding of the relationships between metrics derived from hyperspectral and multispectral sensors to leverage the strengths and mitigate the limitations of each.

Since the inception of the field, spectral vegetation indices (VIs) have been one of the most commonly used tools in the optical remote sensing of vegetation *(Kriegler et al., 1969; Tucker, 1979)*. Over the decades, computational simplicity and ease of use have resulted in a proliferation of VI options (e.g. *Rouse et al. 1974; Huete 1988; Huete et al. 2002; Jiang et al. 2008; Badgley, Field, and Berry 2017*). The growing availability and diminishing cost of decameter *(Claverie et al., 2018; Wulder et al., 2012)* and sub-decameter *(Neigh et al., 2013)* multispectral satellite imagery is making VIs increasingly popular with a yet broader science and applications community. Data from Planet's Dove and SuperDove sensors are of particular interest given their (nominal) daily global coverage and 3 - 4.2 m spatial resolution *(Safyan, 2020)* with constantly improving agronomic applications (*Zhang et al., 2020*).

As recently reviewed by *Swoish et al. (2022), Xue and Su (2017),* and *Zeng et al. (2022)*, popular multispectral VIs vary widely in underlying theoretical basis and conceptual interrelationships. But despite this complexity, independent evaluations of VIs against other approaches for vegetation abundance estimation, particularly using hyperspectral data, are relatively rare. Here, we conducted such a comparative analysis for six popular vegetation indices over a diverse, globally significant agricultural landscape.

Global compilations of multispectral imagery from spectrally diverse landscapes have compared and contrasted interrelationships among vegetation indices through the use of bivariate distributions between pairs of indices *(Small, 2001; Small and Milesi, 2013; Sousa and Small, 2019, 2017)*. Results of these intercomparisons have shown a wide range of interrelationships among vegetation indices often assumed to be interchangeable metrics for the same physical quantity – green vegetation cover at pixel scale. However, these studies have generally relied on decameter (or hectometer) resolution multispectral imagery, which is known to alias spectral signatures and introduce varying degrees of nonlinear spectral mixing within sensor IFOVs at these scales. In addition, global compilations of top-of-atmosphere reflectance spectra generally incorporate a variety of atmospheric and BRDF effects.

We sought to minimize distortions arising from spatial and spectral aliasing, as well as atmospheric and BRDF effects, by using a compilation of 64,000,000 high quality AVIRIS-ng hyperspectra of a diversity of agricultural landscapes in California to quantify characteristics of six widely used broadband vegetation indices. Using spatially and spectrally oversampled reflectances, derived from a state-of-the-art imaging spectrometer, atmospherically corrected with a state-of-the-art radiative transfer model, minimizes the biases inherent in using compilations of broadband data, and allows simultaneous quantitative intercomparison of tens of millions reflectance spectra of a diversity of substrates and vegetation.

Here, we compiled 640,000,000 5 nm, 3-5 m, AVIRIS-ng reflectance spectra from 15 flight lines collected throughout California in 2020. We estimated photosynthetic vegetation fraction ($F_v$) from each AVIRIS spectrum by inverting a three endmember linear spectral mixture model. We then convolved each 425-band AVIRIS reflectance spectrum with the 8-band spectral response of the Planet SuperDove sensor, and used these simulated SuperDove spectra to compute VIs for each spectrum. We quantified the relationship of each VI relative to hyperspectral $F_v$ for the full set of mixed spectra.

In so doing, we addressed the following questions:

1. *How do the linearity and dispersion of each VI distribution vary with photosynthetic vegetation fraction estimated by the spectral mixture model?*

2. *How sensitive is each multispectral VI to substrate background from common agricultural materials like soil, rock, non-photosynthetic vegetation, or plastics?*

3. *How does a traditional parametric similarity metric (Pearson correlation, $\rho$) compare to a popular nonparametric metric (Mutual Information, MI) for such a comparison?*

## Materials and Methods

This study relied upon 15 flight lines collected in late summer of 2020 with the Advanced Visible Infrared Imaging Spectrometer – Next Generation (AVIRIS-ng) instrument. AVIRIS-ng measures radiance from 380 to 2510 nm at 5 nm intervals *(Chapman et al., 2019)*. All flight lines were downloaded from the AVIRIS-ng data portal (https://avirisng.jpl.nasa.gov/dataportal) as orthorectified radiance. Each line was converted to surface reflectance using the Imaging Spectrometer Optimal FITting algorithm (ISOFIT) *(Thompson et al., 2018)*. ISOFIT version 2.9.2 was used, as cloned from https://github.com/isofit/isofit. For all ISOFIT runs, the empirical line method flag was turned on (ELM = 1), with segmentation size = 200.

After atmospheric correction, reflectance data in wavelength ranges 376-421, 1123-1173, 1323-1498, 1774-2024, and 2450-2500 nm (AVIRIS-ng bands 1-10, 150-160, 190-225, 280-330, 415-425) were excluded from statistical analyses due to atmospheric contamination, resulting in retention of 306 of 425 bands. Several 600 x 800 pixel subsets were extracted from each flightline and mosaiced into a single composite image cube of 64,000,000 pixel spectra.

All lines were flown in California between July 24 and September 24, 2020 (Figure 1). Flight altitude for all lines was approximately 10,300 ft (3140 m), resulting in ground sampling distance of roughly 3 m. The lines sampled broad crop and soil diversity within one of the most productive agricultural landscapes on Earth *(CDFA, 2021)*. Flights spanned a broad diversity of soils *(Sousa and Small, 2018)* hosting orchards, vineyards, and row crops, as well as some native and suburban vegetation. Nearly all agriculture in the study area is irrigated and planted/harvested asynchronously, resulting in a wide range of soil exposure, moisture content, growth stage and canopy closure within each line.

Simulated Planet data were computed by convolving AVIRIS-ng surface reflectance with the spectral response of the SuperDove sensor obtained from: https://developers.planet.com/docs/apis/data/sensors/. 6 indices were chosen on the basis of their popularity, as recently reviewed by *(Zeng et al., 2022)*. The indices used in this study are: Difference Vegetation Index (DVI, *(Richardson and Wiegand, 1977)*), Simple Ratio (SR *(Tucker, 1979)*, scaled by 0.1), Normalized Difference Vegetation Index (NDVI, *(Rouse et al., 1974)*) Near-Infrared Reflectance of Vegetation (NIRv *(Badgley et al., 2017)*), Enhanced Vegetation Index (EVI *(Huete et al., 2002)*), and two-band Enhanced Vegetation Index (EVI2 *(Jiang et al., 2008)*). Formulas for each index are given in Supplement Table S2.

Similarity metrics were computed in Python. Pearson correlation coefficients ($\rho$) were computed using NumPy v1.21.2 *(numpy.corrcoef)*. Mutual information (MI) was computed using scikit-learn v0.24.2 *(sklearn.feature_selection.mutual_info_regression)*. MI uses the Kullback-Leibler divergence of the marginal distributions of two variables to quantify the amount of information that can be obtained about one variable by observing the other variable *(Kozachenko and Leonenko, 1987; Kraskov et al., 2004; Ross, 2014; Shannon, 1948)*. Unlike $\rho$, MI does not require parametric statistical assumptions (or even real-valued variables), so should provide a more accurate estimate of the strength of non-Gaussian and nonlinear relationships.

Linear regression estimation for the NIRv:Fv was conducted in the R computing environment (v4.1.1) using base package 'lm'. Negative VI values were excluded when estimating slope.

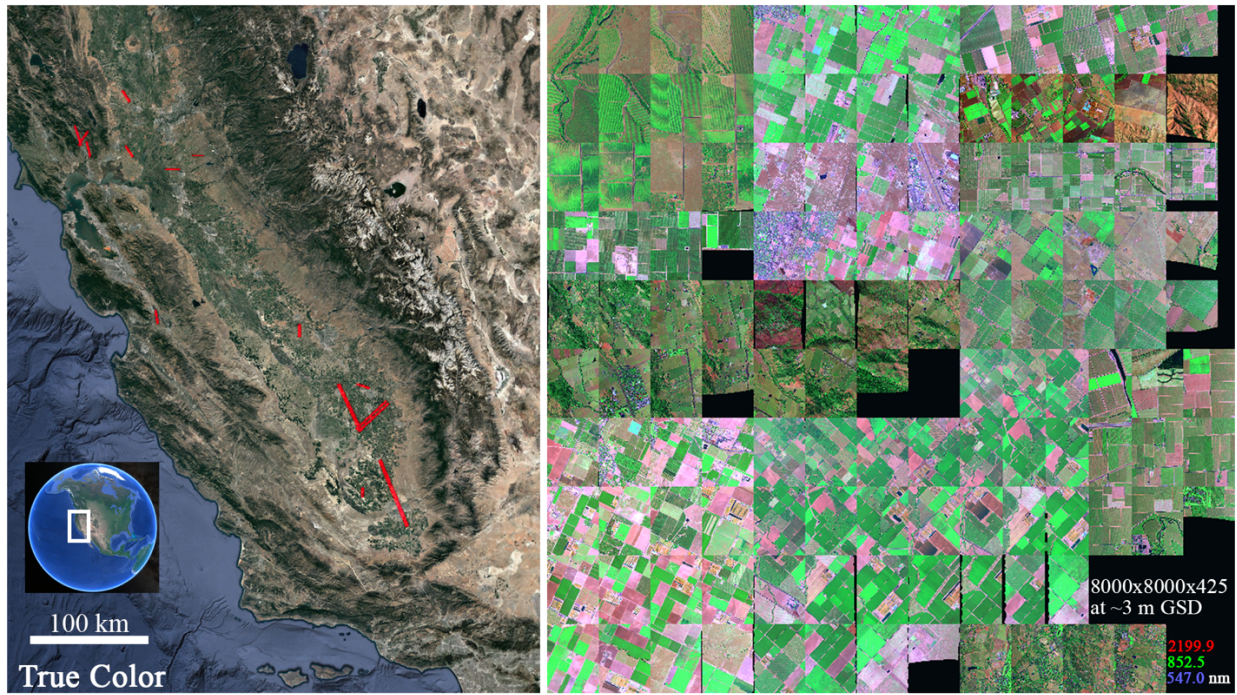

*Figure 1. Index map.* Left: 15 flight lines from the 2020 AVIRIS-ng campaign (red) span broad crop and soil diversity in one of the most productive agricultural regions on earth. Right: False color composite mosaic image compiled from 125 600 x 800 pixel (~2.4 x 3.2 km) subsets. Ground sampling distance for all lines is approximately 3 m. For further information about soil diversity, see (Sousa and Small, 2018) and (NRCS, 2022); for agricultural diversity see (CDFA, 2021).

# Results

The low-order spectral feature space of the AVIRIS-ng reflectance mosaic is shown as orthogonal projections of the low order principal component (PC) distributions in Figure 2. The first three dimensions are found to capture > 97% of the overall variance. The first two dimensions of the spectral feature space are bounded by Substrate, Vegetation, and Dark (S, V, D) spectral endmembers (EMs), with a clear non-photosynthetic vegetation (N) EM emerging in the third dimension. Both 3-EM (SVD) and 4-EM (SVDN) models were inverted and compared. V fraction ($F_v$) estimates from the 3-EM and 4-EM models were found to be strongly interrelated ($\rho > 0.99$; MI = 1.9). $F_v$ from the 3-EM model was used for subsequent analysis.

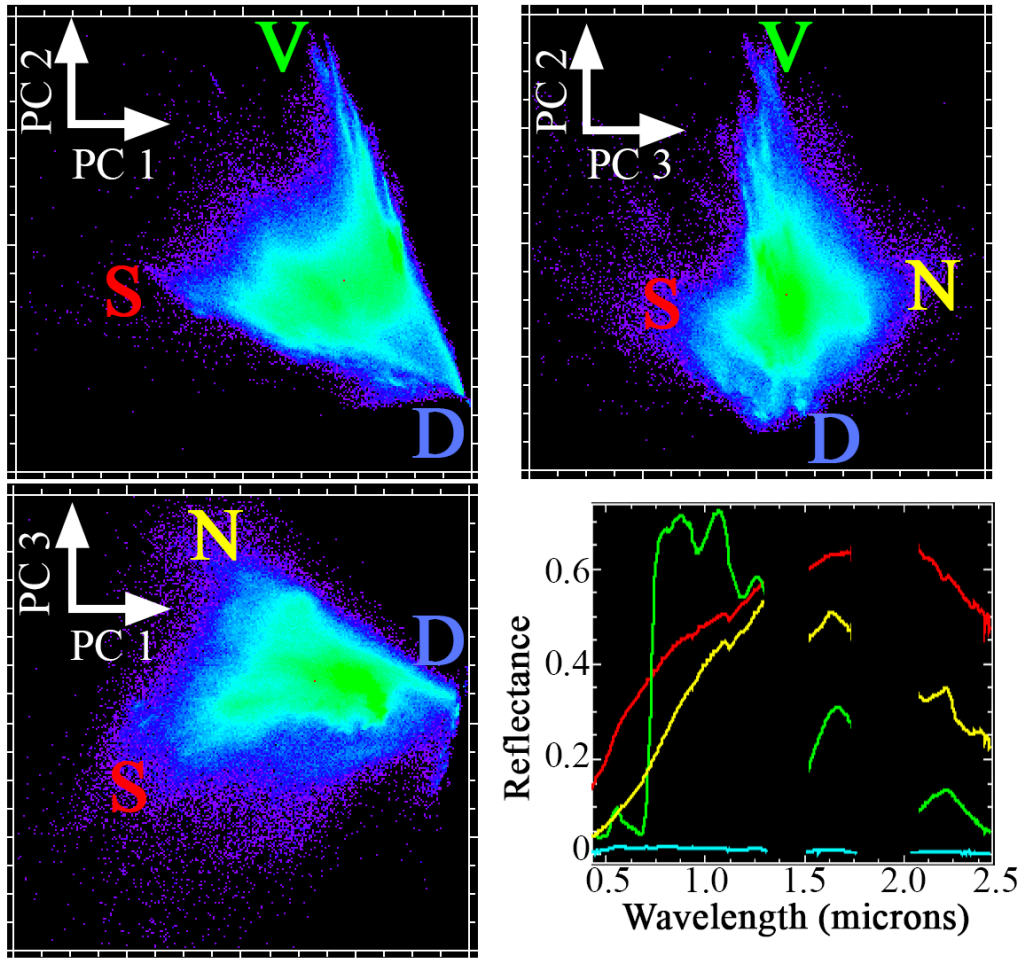

*Figure 2. Spectral feature space and endmembers.* *The first 3 dimensions of the spectral feature space, visualized here, account for over 97% of the variance in the AVIRIS-ng spectra. Substrate, Vegetation and Dark (S, V, D) endmembers bound the first two dimensions of this feature space. Non- photosynthetic vegetation (N) extends the plane of substrates in the third dimension. Compare to previous results from mosaics of AVIRIS-classic and Landsat (Sousa & Small, 2018; Figure 5).*

Bivariate distributions of VIs from SuperDove (y-axis) and $F_v$ from mixture model inversion (x-axis) are shown in Figure 3. DVI and NIRv were very strongly interrelated (DVI:NIRv $\rho > 0.99$; MI > 2.4; Figure S1). DVI and NIRv were also colinear to $F_v$ ($\rho = 0.95$ and MI = 1.4 for both DVI:$F_v$ and NIRv:$F_v$), but substantially biased towards underestimation at higher values (slope =

1.43, intercept = 0.017; regression in Figure S1). Similarly, EVI and EVI2 were also strongly colinear (EVI:EVI2 ρ > 0.99; MI > 2.3; Figure S1). EVI and EVI2 also correlated strongly to $F_v$ (ρ > 0.94; MI > 1.2), showing some overestimation but considerably less bias than DVI or NIRv.

In contrast, NDVI and SR showed considerably more complex relationships to $F_v$. Both ρ (0.84 and 0.81) and MI < 0.7 were much lower than for other indices (Table S3). Consistent with previous comparisons to multispectral mixture models *(Small, 2001; Small and Milesi, 2013; Sousa and Small, 2019, 2017),* NDVI considerably overestimated $F_v$ throughout most of its range, saturating near 0.6. Sensitivity to background reflectance was also especially apparent for NDVI – for instance, spectra with $F_v$ = 0.2 were found to exhibit NDVI values ranging from < 0.05 to > 0.6.

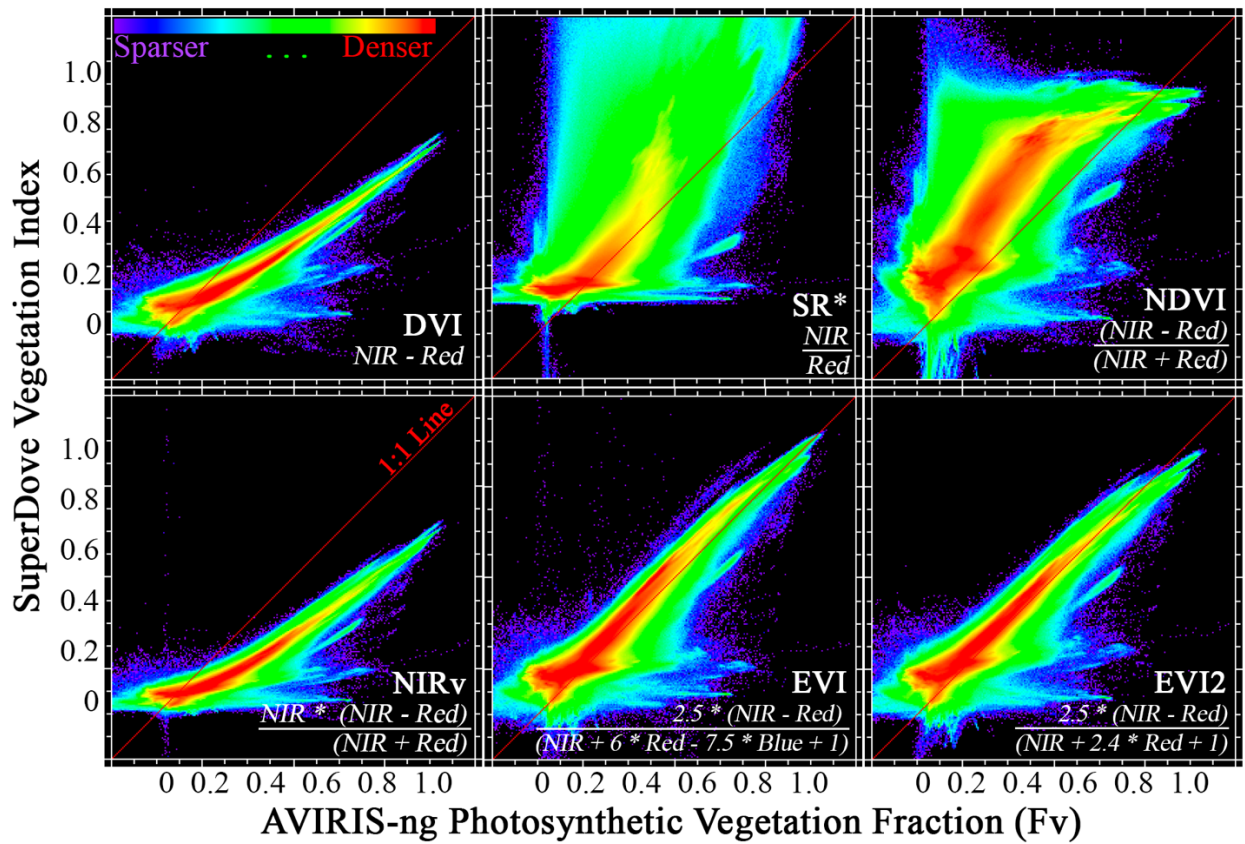

*Figure 3. Bivariate distributions of spectral indices versus vegetation fraction.* 6 commonly used multispectral indices (y-axis) are compared to photosynthetic vegetation fraction ($F_v$) computed directly from hyperspectral AVIRIS-ng reflectances (x-axis). DVI and NIRv are highly correlated with each other (0.99) and with Fv (0.95), but not near the 1:1 line (red). Similarly, EVI and EVI2 are also highly correlated with each other (0.99) and with Fv (0.94 or 0.95). NDVI and SR exhibit substantially reduced correlation to Fv (0.84 and 0.81). Mutual Information (MI) generally agrees with these correlations. MI values (relative to Fv) for DVI, NIRv, EVI, and EVI2 are all 1.35 +/- 0.1 . NDVI and SR MI values are lower, each at 0.69. SR* values are scaled by 0.1, and spectra with values < -0.2 or >1.2 are excluded.

Univariate VI distributions for unvegetated ($F_v$ < 0.05) spectra shown in Figure 4 quantify the sensitivity of each index to substrate and NPV background reflectance. Histograms show only spectra with $F_v$ < 0.05 as estimated by the spectral mixture model. Of the indices, NIRv (yellow) exhibited the least sensitivity to background reflectance, with a mode near 0.05 and standard deviation (σ) = 0.018. DVI (red) demonstrated slightly more sensitivity, with a greater mode (0.08) and dispersion (σ = 0.029). EVI (magenta) and EVI2 (white) performed similarly to each other, with greater modes (0.12) and dispersions (σ = 0.038). Interestingly, SR (green) performed similarly to EVI and EVI2, with a similar mode (0.14) and lower dispersion (σ = 0.018). Consistent with the bivariate distributions showed in Figure 3, NDVI (cyan) demonstrated the most severe sensitivity to substrate background, with a modal value for unvegetated spectra above 0.2, higher dispersion (σ = 0.083), and an upper tail extending beyond 0.5.

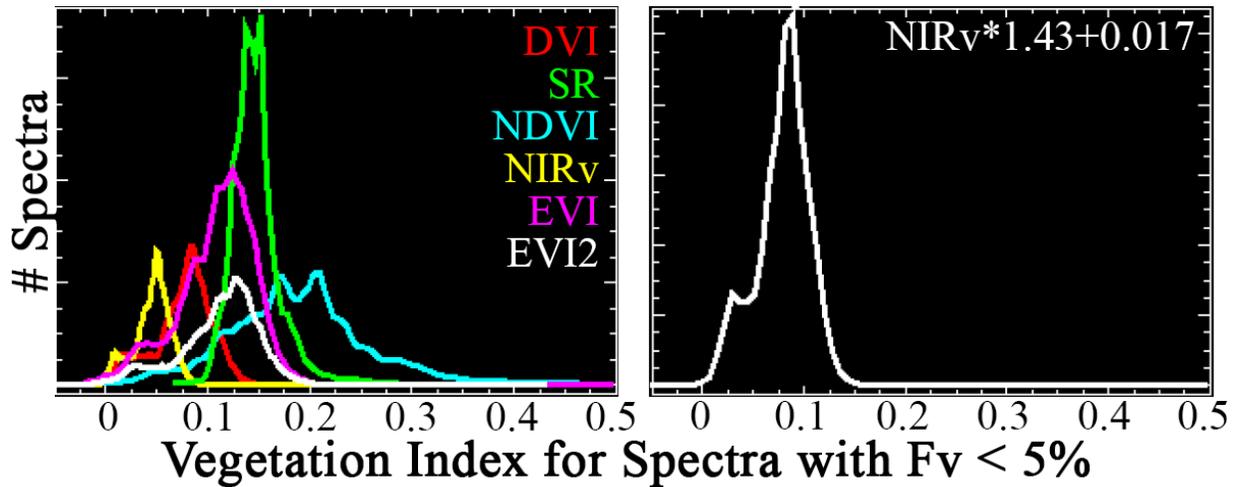

**Figure 4. Spectral index distributions for unvegetated spectra.** *Left: Histograms show VI values for all spectra with < 5% photosynthetic vegetation cover, as estimated from inversion of the AVIRIS-ng spectral mixture model. NIRv values are closest to zero with nearly all values < 0.1 and a mode near 0.05. DVI is characterized by a mode near 0.08 and greater dispersion. EVI and EVI2 both show modes near 0.12. SR modal value is near EVI, but with less dispersion. NDVI shows, by far, the highest mode (> 0.2) and largest dispersion (NDVI for some unvegetated spectra as large as 0.5). Right: Histogram of NIRv values after linear regression is applied. For the distribution of regressed NIRv, mean = 0.08 and standard deviation = 0.026.*

Both metrics of statistical similarity are found to yield similar overall patterns. Pairwise values (Tables S3 and S4 and Figure S1) are approximated by a monotonic, log-linear relationship. An important exception is found for the SR:NDVI relationship which deviates markedly towards high MI values (MI > 10) without a coincident increase in ρ – further discussed below.

## Discussion

### Vegetation index intercomparison

A principal finding of this analysis is the observed similarity of two pairs of vegetation indices: EVI with EVI2, and DVI with NIRv. The similarity between EVI and EVI2 confirms the potential for EVI2 to serve as a representative proxy for EVI when a blue band is unavailable. The strong correlation and low bias between both these indices and $F_v$ supports their use for the estimation of vegetation abundance in cases where SWIR bands are not available. Likewise, the similarity of DVI and NIRv, at least for this diverse agricultural mosaic, is striking (ρ > 0.99; MI = 2.45, 99% of |DVI-NIRv| < 0.07). The minimal sensitivity to background reflectance trades off with overall underestimation of $F_v$. While this bias can be corrected using a simple linear model, the correction increases sensitivity to background reflectance for sparse vegetation.

In contrast, NDVI and SR deviate from all 4 other VIs in important ways. NDVI in particular is challenged by saturation at mid-to-upper values and sensitivity to background reflectance at mid-to-lower values, both of which are clearly more severe than for the other VIs. These findings extend previous results based on datasets collected at coarser spatial and spectral resolution *(Elmore et al., 2000; Small, 2001; Small and Milesi, 2013; Sousa and Small, 2019, 2017)* to 3 m imaging spectroscopy, and agree with a long literature (e.g., (*Gitelson, 2004; Myneni et al., 1995)*, including a recent multiscale results correlating to cover crop biomass (Swoish et al., 2022). Concordance with these studies suggests it is reasonable to expect the results of this analysis – including the striking heteroskedasticity, nonlinearity, and background reflectance sensitivity of NDVI and SR – to generalize to other current and future satellite (and airborne) multispectral sensors like those on the WorldView, Pleiades, Landsat and Sentinel-2 platforms.

## Similarity metrics

Another finding of this analysis is the correspondence between parametric (Pearson correlation coefficient, ρ) and nonparametric (Mutual Information; MI) methods of quantifying statistical similarity and information content. This consistency strengthens the findings of the VI intercomparison, as both methods show strong relationships between [$F_v$] and [DVI, NIRv, EVI, and EVI2] but much weaker relationships between [$F_v$] and [SR, NDVI]. Reduced MI values for NDVI suggest that the limitations of the index are yet more pernicious than simply being due to a nonlinear, heteroskedastic relationship (which should be better captured by MI). This is supported by the broader dispersion of the NDVI distribution than is evident in DVI, NIRv, EVI, or EVI2 in Figure 3. Note also that NDVI and SR have, by far, the largest value of MI (11.33) of any pair of indices, consistent with their demonstrable nonlinear relationship (Figure S1, lower right). This is due to a functional dependence between the two indices, further explored in Figure S1 and Analytical Exercise S1. Such behavior is notably absent for the NDVI:Fv and SR:Fv relationships, which do not markedly deviate above ρ:MI loglinearity, suggesting that even nonlinear regression is unlikely to be effective in capturing this variability.

## Dimensionality and spectral endmembers

Another important finding of this analysis is the broad agreement with previous studies in terms of both spectral dimensionality *(Boardman and Green, 2000; Cawse-Nicholson et al., 2019; Sousa et al., 2022; Sousa and Small, 2018; Thompson et al., 2017)*, and generality of S,V,D endmember spectra *(Small, 2004; Small and Milesi, 2013; Sousa et al., 2022; Sousa and Small, 2018, 2017)*. The prominence of the NPV EM is more evident than in some prior studies, which may be due to the focus of the study on late summer agricultural landscapes with spatially extensive monocultures of dense vegetation at various stages in its life cycle. In this study, the clear differentiation of S and N EMs along PC3 is especially interesting given the dichotomy of 1) broad similarity of their spectral continua, versus 2) with important differences in cellulose/lignin absorption features at SWIR wavelengths *(Roberts et al., 1993)*.

## Limitations and future work

This study was designed to compare the theoretical performance of various multispectral vegetation indices against the results from a hyperspectral mixture model. Deriving simulated Planet SuperDove reflectance directly from atmospherically corrected AVIRIS-ng spectra controls for differences in sun-sensor geometry, sensor-to-sensor radiometric miscalibration, image coregistration, and atmospheric correction that would bias a comparison of independently collected observations. We note that radiometric and BRDF differences may be especially significant for Planet data, given the known limitations of the Dove/SuperDove fleet – and constantly evolving remedies *(Leach et al., 2019; Vanhellemont, 2019; Wegmueller et al., 2021)*. Additionally, by convolving AVIRIS-ng reflectance spectra to SuperDove spectral response functions after applying ISOFIT, it is likely that the simulated SuperDove spectra exhibit substantially more accurate atmospheric correction than would be possible using SuperDove radiance spectra directly. This is most likely to impact results for EVI, given its use of the visible blue band. Moving forward, a comparative analysis of EVI and EVI2 computed from a comparably diverse compilation of SuperDove-observed reflectance spectra might provide a constraint on effectiveness of atmospheric correction and could be an interesting avenue for future work.

## Conclusions

Using a compilation of 125 subsets from 15 airborne imaging spectroscopy flight lines spanning a diverse agricultural mosaic, we simulated 64,000,000 multispectral SuperDove reflectance spectra and used them to compute 6 commonly used vegetation indices. We then compared these indices against photosynthetic vegetation fraction estimated from inversion of a linear mixture model applied to the original 5 nm hyperspectral data ($F_v$). We quantified complementary aspects of the bivariate distributions using both parametric ($\rho$) and nonparametric (MI) metrics. DVI and NIRv were strongly colinear ($\rho > 0.99$; MI = 2.45) and approximately linearly related to $F_v$ ($\rho > 0.95$; MI > 1.4), with minimal sensitivity to substrate background reflectance (0 to 0.1 for unvegetated spectra) but substantial deviation from 1:1 (DVI and NIRv values of 0.6 for $F_v$ values of 1.0). EVI and EVI2 also performed similarly to each other ($\rho > 0.99$; MI = 2.3) and were also approximately linear to $F_v$ ($\rho = 0.95$; MI > 1.2) – but with both more prominent impact of background reflectance (0 to 0.2 for unvegetated spectra) and greatly reduced deviation from 1:1. Of the indices, NDVI and SR exhibited by far the weakest relationships to $F_v$ ($\rho = 0.84$ and 0.81; MI = 0.69). NDVI in particular showed severe saturation effects and sensitivity to soil background reflectance. Comparison of parametric ($\rho$) and nonparametric (MI) similarity statistics yielded a roughly log-linear relationship, with a notable exception supporting a nonlinear analytical relationship between NDVI and SR.

Taken together, this analysis suggests that for agronomic applications:

1. EVI2 can likely serve as a proxy for EVI in situations where a blue band is not available, and both can serve as proxies for $F_v$ in situations where SWIR bands are not available.

2. DVI and NIRv yield very similar values ($\rho > 0.99$, 99% of |DVI-NIRv| < 0.07).

3. DVI and NIRv show low sensitivity to substrate background reflectance, but systematically underestimate $F_v$. Linear bias correction reduces deviation from 1:1 but increases substrate background effects.

4. Bivariate distributions of ρ and MI can usefully contextualize nonlinear VI relationships.

5. As metrics of subpixel vegetation abundance, NDVI and SR exhibit severe challenges in nonlinearity, heteroskedasticity, and sensitivity to substrate background reflectance.

# Acknowledgements

D.S. was supported by the USDA NIFA Sustainable Agroecosystems Program under Grant # 2022-67019-36397; the NASA Land-Cover/Land-Use Change Program under Grant # LCLUC21_2-0025; and the NASA Remote Sensing of Water Quality Program under Grant # 80NSSC22K0907. C.S. was supported by the endowment of the Lamont Doherty Earth Observatory of Columbia University. The authors thank the National Center for Ecological Analysis and Synthesis (NCEAS) for providing computational resources that supported this work.

# Declaration of Interest Statement

The authors report there are no competing interests to declare.

# Supplement

**Table S1. AVIRIS-ng flightlines used for this analysis.** Both short name (this analysis) and full flightline ID (archived on the JPL database) are provided. 15 lines total.

| Short Name | Flightline ID |
|---|---|
| Gilroy | ang20200918t232303 |
| Kern_1 | ang20200724t191126 |
| Kern_2 | ang20200924t213537 |
| Kings | ang20200924t200728 |
| Lodi_1 | ang20200907t203701 |
| Lodi_2 | ang20200918t210935 |
| MaderaFresno | ang20200924t203044 |
| Napa_1 | ang20200918t215728 |
| Napa_2 | ang20200918t220357 |
| Napa_3 | ang20200918t221604 |
| Solano | ang20200918t204940 |
| Tulare_1 | ang20200903t201645 |
| Tulare_2 | ang20200903t203648 |
| TulareKings | ang20200924t193402 |
| Yolo | ang20200918t203620 |

**Table S2. Spectral index formulas.** All indices computed from simulated SuperDove spectra.

| Index Name | Acronym | Formula |
|---|---|---|
| Difference vegetation (veg.) index | DVI | NIR – Red |
| Normalized difference veg. index | NDVI | (NIR – Red) / (NIR + Red) |
| Near-infrared reflectance of veg. | NIRv | NIR * (NIR – Red) / (NIR + Red) |
| Simple Ratio | SR | NIR / Red |
| Enhanced veg. index | EVI | 2.5 * (NIR – Red) / (NIR + 6 * Red – 7.5 * Blue + 1) |
| 2-band Enhanced veg. index | EVI2 | 2.5 * (NIR – Red) / (NIR + 2.4 * Red + 1) |

**Table S3. Mutual information matrix.**

*Mutual Information (MI)*

|       | Fv    | DVI   | NDVI  | NIRv  | SR    | EVI   | EVI2  |
|---|---|---|---|---|---|---|---|
| Fv    | 12.01 | 1.44  | 0.69  | 1.41  | 0.69  | 1.25  | 1.34  |
| DVI   | 1.44  | 12.01 | 0.77  | 2.45  | 0.77  | 1.60  | 1.80  |
| NDVI  | 0.69  | 0.77  | 12.01 | 0.98  | 11.34 | 1.20  | 1.25  |
| NIRv  | 1.41  | 2.45  | 0.98  | 12.01 | 0.98  | 2.01  | 2.77  |
| SR    | 0.69  | 0.77  | 11.33 | 0.98  | 12.01 | 1.20  | 1.25  |
| EVI   | 1.25  | 1.60  | 1.20  | 2.01  | 1.20  | 12.01 | 2.30  |
| EVI2  | 1.34  | 1.80  | 1.25  | 2.77  | 1.25  | 2.30  | 12.01 |

**Table S4. Correlation matrix.**

*Pearson Correlation (ρ)*

|      | Fv    | DVI   | NDVI  | NIRv  | SR    | EVI   | EVI2  |
|-----:|-------|-------|-------|-------|-------|-------|-------|
| **Fv**   | 1.000 | 0.950 | 0.837 | 0.949 | 0.806 | 0.940 | 0.949 |
| **DVI**  | 0.950 | 1.000 | 0.826 | 0.992 | 0.818 | 0.973 | 0.982 |
| **NDVI** | 0.837 | 0.826 | 1.000 | 0.860 | 0.913 | 0.904 | 0.910 |
| **NIRv** | 0.949 | 0.992 | 0.860 | 1.000 | 0.873 | 0.986 | 0.990 |
| **SR**   | 0.806 | 0.818 | 0.913 | 0.873 | 1.000 | 0.886 | 0.882 |
| **EVI**  | 0.940 | 0.973 | 0.904 | 0.986 | 0.886 | 1.000 | 0.993 |
| **EVI2** | 0.949 | 0.982 | 0.910 | 0.990 | 0.882 | 0.993 | 1.000 |

**Figure S1. Additional VI relationships.** Upper left: DVI and NIRv are highly correlated (ρ >0.99), but DVI gives slightly higher values (mean difference 4.0%, standard deviation 1.2%). Upper right: EVI and EVI2 are also highly correlated (ρ > 0.99), with a much smaller average difference (mean = 0.1%) but greater dispersion (standard deviation = 2.2%). Lower left: Regressing NIRv against Fv greatly reduces underestimation but increases the sensitivity to substrate background reflectance (note negative values excluded in regression). Lower right: The bivariate distribution of NDVI and SR gives a strikingly tight curvilinear relationship. An algebraic explanation for this is explored in Analytical Exercise S1.

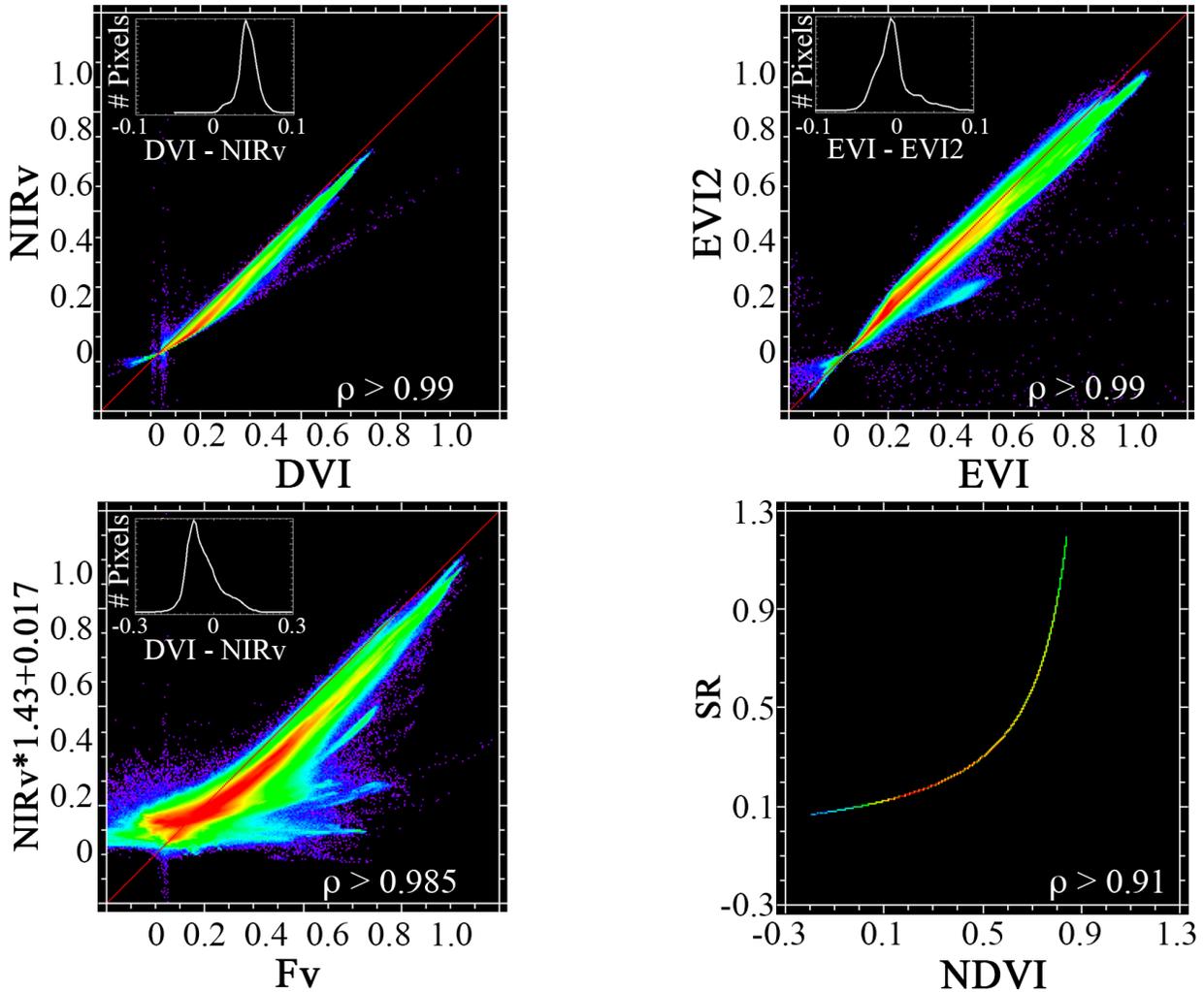

**Figure S2. Parametric versus nonparametric statistics.** Pearson correlation coefficient (ρ) is roughly loglinear with Mutual Information (MI) for these distributions. The strong nonlinear analytic NDVI:SR relationship (lower right on Figure S1) occurs as an outlier deviating well above the linear relation (ρ = 0.91, $\log_{10}$(MI) > 1). This demonstrates the efficacy of MI in quantifying nonlinear relationships. The lack of similarly elevated MI values for NDVI:$F_v$ and SR:$F_v$ provides further evidence that the greater dispersion and heteroskedasticity of these indexes would be challenging to incorporate effectively into even a nonlinear regression. The 7 identical outliers (ρ =1.0, MI = 12.01) upper right correspond to self-information of each distribution with itself.

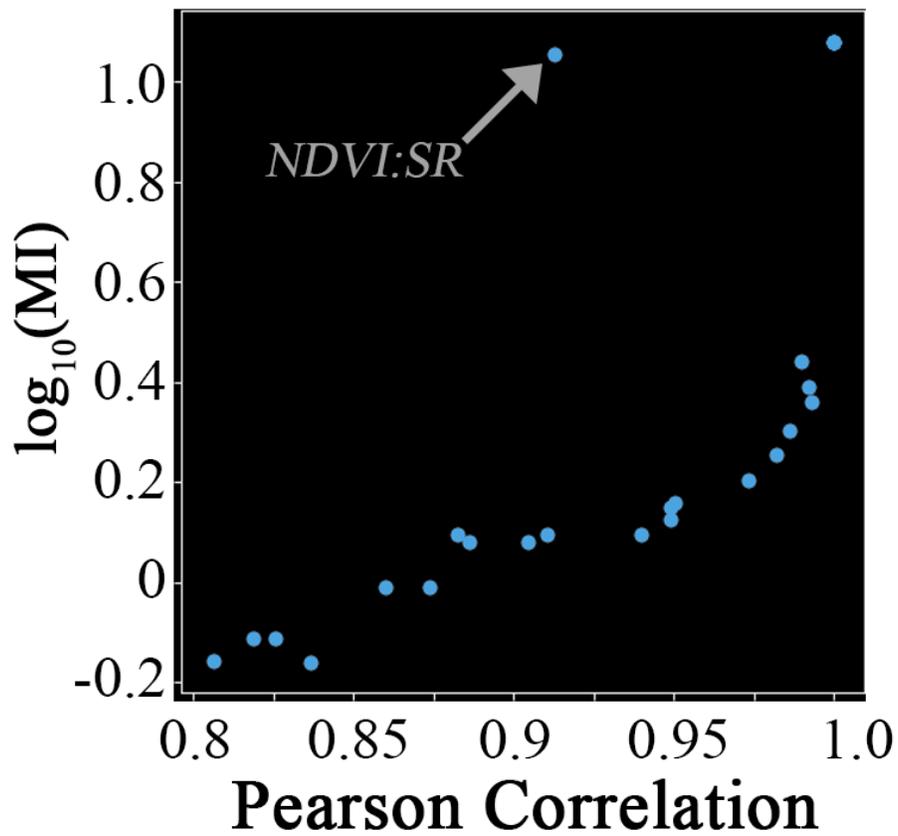

**Analytical Exercise S1.** An exploration of the relationship between SR and NDVI.

Begin with the formula for SR:

$$SR = \frac{NIR}{Red}$$

Rearrange terms:

$$Red = \frac{NIR}{SR}$$

Now examine the formula for NDVI:

$$NDVI = \frac{NIR - Red}{NIR + Red}$$

Substitute for Red:

$$NDVI = \frac{NIR - \frac{NIR}{SR}}{NIR + \frac{NIR}{SR}}$$

Multiply by 1:

$$NDVI = \frac{NIR - \frac{NIR}{SR}}{NIR + \frac{NIR}{SR}} \times \frac{SR}{SR}$$

$$NDVI = \frac{(SR \times NIR) - NIR}{(SR \times NIR) + NIR}$$

Factor:

$$NDVI = \frac{NIR \times (SR - 1)}{NIR \times (SR + 1)}$$

Simplify:

$$NDVI = \frac{SR - 1}{SR + 1}$$

The relationship between SR and NDVI can thus be described by a simple rational function of the form:

$$y = \frac{x - 1}{x + 1}$$

This explains the curvilinear shape of the lower right plot in Figure S1, as well as the notably elevated MI score for this pair of vegetation indices.